# Wyner wiretap-like encoding scheme for cyber-physical systems

*Walter Lucia*[1] ✉, *Amr Youssef*[1]
[1]*Concordia Institute for Information Systems Engineering (CIISE), Concordia University, Montreal, Canada*
✉ *E-mail: walter.lucia@concordia.ca*

**Abstract:** In this study, the authors consider the problem of exchanging secret messages in cyber-physical systems (CPSs) without resorting to cryptographic solutions. In particular, they consider a CPS where the networked controller wants to send a secrete message to the plant. They show that such a problem can be solved by exploiting a Wyner wiretap-like encoding scheme taking advantage of the closed-loop operations typical of feedback control systems. Specifically, by resorting to the control concept of one-step reachable sets, they first show that a wiretap-like encoding scheme exists whenever there is an asymmetry in the plant model knowledge available to control system (the defender) and to the eavesdropper. The effectiveness of the proposed scheme is confirmed by means of a numerical example. Finally, they conclude the study by presenting open design challenges that can be addressed by the research community to improve, in different directions, the secrete message exchange problem in CPSs.

## 1 Introduction

Cyber-physical systems (CPSs) refer to 'smart' co-engineered interacting networks of physical and computational components [1]. Examples of CPSs include smart grids, aviation systems, nuclear power plants, water supply systems, and industrial control systems. While CPSs provide richer functionality, efficiency, autonomy, and reliability compared to manually controlled and loosely coupled systems, they may also create inherent security vulnerabilities [2–4]. Ensuring the security of current and emerging CPSs must take into consideration the unique challenges present in their environment [5].

Over the past decade, different control schemes have been proposed for the detection of false data injection attacks in CPSs [6–12]. Of particular interest are the classes of solutions against stealthy attacks [13] whose implementations require the exchange of a secret message between the plant and the control centre, see e.g. the moving target solutions and sensor coding schemes developed in [14–18]. Traditionally, this key agreement step is achieved through the use of cryptographic protocols, which can be classified as symmetric key protocols or public key protocols [19]. Such cryptographic approaches might not always be suited for CPSs. For example, symmetric key-based solutions assume the existence of a pre-shared key. However, the compromise of such long-term keys usually leads to compromising the security of the whole system. On the other hand, public key protocols might be computationally prohibitive in some CPSs' environments and also require the establishment of a public key infrastructure [19], including the support of a key revocation mechanism (e.g. see [20]), which might be hard to deploy in many CPSs' applications.

Another approach to secure communications without relying on classical cryptographic approaches was introduced by Wyner [21], who presented the wiretap channel model. This model utilises the role of noise, which is a natural characteristic in any communications system, to achieve secure communications.

In Wyner's model, we assume that the eavesdropper observes a noisy version of the signal available at the receiver. The defender's objective is then to design an encoding scheme such that, for sufficiently large codeword length, the equivocation rate [22] of the transmitted message given the signal received by the adversary is arbitrarily close to the entropy rate of the message [23].

In this paper, we consider a networked control system, where the (networked) controller wants to send a secrete message to the plant without resorting to cryptographic solutions. Our control theoretic solution utilises the asymmetry in the plant model knowledge available to control system designer/operator (the defender) and to the eavesdropper. This asymmetry can be exploited to play a role similar to the one performed by the noise in Wyner's wiretap channel, and hence allows us to design a wiretap-like encoding scheme for CPSs.

The rest of the paper is organised as follows. In Section 2, we briefly review related works. The preliminaries and our problem formulation are presented in Section 3. The proposed control theoretic encoding scheme, which allows us to transfer secret messages in CPSs without relying on classical cryptographic solutions, is developed in Section 4 and its effectiveness is verified through simulation in Section 5. Finally, we conclude the paper in Section 6 where we also present some open research directions.

## 2 Related work

Random noise is an intrinsic element of almost all physical communication channels. In an effort to understand the role of noise in the context of secure communications, in a landing marking paper [21], Wyner introduced the wiretap channel. Wyner's wiretap channel models a legitimate transmitter (Alice) and a receiver (Bob) communicating over a discrete memoryless channel, referred to as the main channel, in the presence of a passive wiretapper who only listens to the transmitted signal through a second channel, referred to as the wiretapper channel. The goal is to design a coding scheme that makes it possible for Alice to communicate both reliably and securely. Reliability is measured in terms of Bob's probability of error in recovering the message, while security is measured in terms of the mutual information [22] between the message and Eve's observations. Wyner showed that the situation is characterised by a single constant $C_s$, called the secrecy capacity, which has the following meaning: for all $\epsilon > 0$, there exist coding schemes of rate $R \geq C_s - \epsilon$ which asymptotically achieve both the reliability and the security objectives.

The fundamental role of noise has also been utilised in the design of secret-key agreement schemes, without relying on cryptographic methods (e.g. see Chapter 4 in [23]). In this approach, we assume that the legitimate parties and the eavesdropper observe the realisations of correlated random variables and that the legitimate parties attempt to agree on a secret key unknown to the eavesdropper. To isolate the role played by noise, the legitimate parties are assumed to be able to distil their key by communicating over a two-way, public, noiseless, and authenticated channel. The relevant metric, referred to as the key secrecy capacity, is the number of secret-key bits that can be distilled per observation of the correlated random variables.

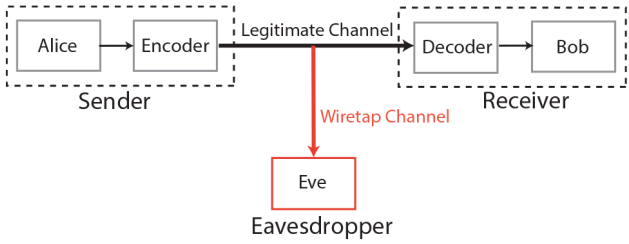

**Fig. 1** *Wyner wiretap channel*

The concept of wiretap channel [21] and secrecy capacity (also see [24]) has prompted significant research in physical layer security of wireless communication systems [23] over the past years. Such approaches have been extended to secure communications in CPSs. Burg *et al.* [25] provides a summary of communication principles from the perspective of the connectivity needs of CPSs. The gap between the security features in the communication standards used in CPSs and IoT and their actual vulnerabilities are pointed out. The authors concluded their paper by emphasising the need for a more in-depth study of the security issues across all the protocol layers, including both logical layer security and physical layer security. Atat *et al.* [26] discussed the main challenges that cellular providers will face as the massive number of CPS devices attempt to access the cellular spectrum. They also study how to protect the device-to-device links from eavesdropping through the use of an aggressive sensing technique.

Rawat *et al.* [27] analysed outage probability for secrecy rate in multiple-input multiple-output wireless systems in the presence of eavesdroppers and jammers for CPS devices. However, similar to other existing literature, the authors followed a purely information theoretic approach that does not utilise the dynamics of the underlying CPSs.

Recently, some works have utilised control-theoretic approaches to show the existence of covert channels [28, 29] in CPSs. Such channels can be abused by attackers to exfiltrate sensitive information, such as the proprietary gains or the thresholds used in the controller or to coordinate for CPS attacks. For example, Krishnamurthy *et al.* [30] proposed using the analogue emissions of physical instrumentation such as actuators, and sensors in CPS to send or leak information without impacting the CPS process characteristics. As a proof of concept, they demonstrated that a malware can use the acoustic emissions of a motor controlling a valve in a feedback control loop as a covert channel without affecting the stability, performance, and signal characteristics of the closed-loop process. Garcia *et al.* [31] presented a covert channel that leverages physical substrates, such as line loads within a power system, to transmit information between compromised controllers. Such channel can be used, by the compromised controllers, in a coordinated attack scenarios by manipulating relays to modify the power network's topology without the use of any explicit communication channels (e.g. power line communications) in order to evade intrusion detection. In [32], Herzberg and Kfir showed how a corrupt actuator in one CPS zone can send covert information to a sensor in a different zone, breaking the isolation. This may allow an attack where the actuator is intentionally malfunctioning, and the sensor is intentionally masking the malfunction. In [33], the same authors presented another covert channel from a covertly-transmitting sensor to an actuator, interacting only indirectly, via a benign threshold-based controller. The covert traffic is encoded within the output noise of the covertly-transmitting sensor, whose distribution is indistinguishable from that of a benign sensor. Finally, in [34], Ying *et al.* utilised covert channels to build an effective defensive technique that facilitates transmitter authentication via a trusted monitor node for automotive applications. In particular, the authors presented transmitter authentication for controller area network (TACAN), which provides secure authentication of electronic control units (ECUs) by exploiting the covert channels without introducing CAN protocol modifications or traffic overheads. TACAN exploits different covert channels for ECU authentication including inter-arrival time of CAN messages, and the clock offsets of CAN messages. It also conceals authentication messages into the least significant bits of normal CAN data.

## 3 Preliminaries and problem formulation

For basic definitions related to information theory (e.g. entropy, mutual information, and equivocation), we refer the reader to [22]. A good introduction to the principles of physical layer security can be found in [23].

In what follows, we denote with $\mathbb{R}$, $\mathbb{R}^{n_v}$, and $\mathbb{Z}_+ = \{0, 1, \ldots\}$, the sets of real numbers, real-values column vectors of dimension $n_v > 0$, and non-negative integer numbers, respectively. Moreover, given a variable $v$, $v(k)$ denotes the value of $v$ at the discrete sampling time instant $k \in \mathbb{Z}_+$.

*Definition 1:* Given two sets $\mathcal{R} \subset \mathbb{R}^n$ and $\mathcal{Q} \subset \mathbb{R}^n$, the set-difference $\mathcal{R} \setminus \mathcal{Q}$ and the Minkowski sum $\mathcal{R} \oplus \mathcal{Q}$ are defined as follows [35]:

$$\mathcal{R} \setminus \mathcal{Q} := \{\mathbf{r} \in \mathbb{R}^n : \mathbf{r} \in \mathcal{R}, \mathbf{r} \notin \mathcal{Q}\} \quad (1)$$

$$\mathcal{R} \oplus \mathcal{Q} := \{\mathbf{r} + \mathbf{q} \in \mathbb{R}^n : \mathbf{r} \in \mathcal{R}, \mathbf{q} \in \mathcal{Q}\} \quad (2)$$

*Definition 2:* Consider the discrete-time system

$$\mathbf{x}(k+1) = f(\mathbf{x}(k), \mathbf{u}(k), \mathbf{d}(k)) \quad (3)$$

where $\mathbf{x}(k) \in \mathbb{R}^n$ and $\mathbf{u}(k) \in \mathbb{R}^m$ are the state and input vectors, respectively, $\mathbf{d}(k) \in \mathbb{R}^d \in \mathcal{D}$ is a bounded exogenous disturbance with $\mathcal{D}$, a compact set with $\mathbf{0}_d \in \mathcal{D}$, and $f(\cdot, \cdot, \cdot)$ is a function describing the system dynamics.

Let $\mathbf{x} \in \mathbb{R}^n$ be the current state of the system, then the set of states one-step reachable using the control input $\mathbf{u} \in \mathbb{R}^m$ is defined as follows [36]:

$$\text{Reach}(\mathbf{x}, \mathbf{u}) = \{\mathbf{x}^+ \in \mathbb{R}^n : \exists \underline{\mathbf{d}} \in \mathcal{D} \text{ s.t. } \mathbf{x}^+ = f(\mathbf{x}, \mathbf{u}, \underline{\mathbf{d}})\} \quad (4)$$

### 3.1 Wyner wiretap channel

Consider the situation depicted in Fig. 1, where Alice (sender) wants to exchange a secrete message with Bob (receiver) using a legitimate memoryless communication channel and Eve (eavesdropper) wants to intercept the secret by establishing another memoryless channel known as the wiretap channel [21]. Moreover, assume that the sender and receiver can transmit messages using any encoding/decoding scheme but that these operations are also known to the eavesdropper.

*Proposition 1:* If the equivocation of the encoded data seen by the eavesdropper is greater or equal to uncertainty of the data source, then there exists an encoding scheme allowing Alice to send a secrete message to Bob with approximately perfect secrecy [21].

### 3.2 Networked control system

Consider the networked CPSs shown in Fig. 2, where the plant dynamics are described by (3).

*Assumption 1:* We assume that the exact dynamical model (3) is unknown and that, for control purposes, an appropriate system identification procedure [37] has been offline used to approximate the system dynamics. Therefore, without loss of generality, we assume that the following uncertain model is used by the controller:

$$\mathbf{x}(k+1) = f_c(\mathbf{x}(k), \mathbf{u}(k)) + \mathbf{d}_c(k), \mathbf{d}_c(k) \in \mathcal{D}_c \quad (5)$$

where $\mathcal{D}_c \subset \mathbb{R}^d$ is a compact subset with $\mathbf{0}_d \in \mathcal{D}_c$. Moreover, $f_c(\mathbf{x}(k), \mathbf{u}(k)) + \mathbf{d}(k)$ is a function satisfying the inclusion condition

$$\forall k \in \mathbb{Z}_+ \exists \underline{\mathbf{d}}_c \in \mathcal{D}_c : f(\mathbf{x}(k), \mathbf{u}(k), \mathbf{d}(k)) = f_c(\mathbf{x}(k), \mathbf{u}(k)) + \underline{\mathbf{d}}_c \quad (6)$$

which guarantees that the system dynamics (3) are contained into the identified uncertain model (5).

Moreover, given (6), we assume that the controller logic is generically described by the following state-feedback control law:

$$\mathbf{u}(k) = \phi(\mathbf{x}(k)) \quad (7)$$

### 3.3 Attacker model

*Assumption 2:* We assume that the communication channels between the controller and the plant can be subject to eavesdropping attacks capable of reading all the transmitted data. Moreover, the attacker, exploiting the available resources, is capable of obtaining the following plant uncertain model:

$$\mathbf{x}(k+1) = f_e(\mathbf{x}(k), \mathbf{u}(k), \mathbf{d}_e(k)), \mathbf{d}_e(k) \in \mathcal{D}_{f_e} \quad (8)$$

with $\mathcal{D}_{f_e} \subset \mathbb{R}^d$ a compact subset and $f_e(\mathbf{x}(k), \mathbf{u}(k), \mathbf{d}_e(k))$ a function such that

$$\forall k \in \mathbb{Z}_+ \exists \underline{\mathbf{d}}_e \in \mathcal{D}_{f_e} : f(\mathbf{x}(k), \mathbf{u}(k)) = f_e(\mathbf{x}(k), \mathbf{u}(k), \underline{\mathbf{d}}_e) \quad (9)$$

Hereafter, for the sake of clarity, we describe (8) by means of the following model:

$$\mathbf{x}(k+1) = f_c(\mathbf{x}(k), \mathbf{u}(k)) + \mathbf{d}_e(k), \mathbf{d}_e(k) \in \mathcal{D}_e \quad (10)$$

with $\mathcal{D}_e \subset \mathbb{R}^d$ a compact subset with $\mathbf{0}_d \in \mathcal{D}_e$, and such that

$$\mathcal{D}_c \subset \mathcal{D}_e \quad (11)$$

*Remark 1:* Note that (11) assures that the model used by the controller (5) is more accurate than the model available to the attacker (10). The latter finds justification by the fact that the defender can actually design the system identification process by adequately selecting sufficiently exciting input signals [38]. On the other hand, the attacker cannot design the input signals and can perform system identification only using the data collected during the online closed-loop operations.

### 3.4 Problem formulation

The objectives of the paper can be formally stated as follows: given the networked control system in Fig. 2, the plant dynamics (3), and Assumptions 1 and 2:

- Demonstrate the existence of a Wyner wiretap-like encoding scheme for the secrete exchange of messages between the controller (sender) and the plant (receiver).
- Design a proof-of-concept encoding scheme for transferring secret messages from the controller to the plant.

## 4 Wyner wiretap-like channel in CPS and the proposed encoding scheme

In this section, we first show the existence of the Wyner wiretap-like channel in CPSs. Then, as a proof-of-concept, an encoding scheme is designed, and the resulting algorithm is presented.

### 4.1 Wyner wiretap-like channel in CPS

Consider a scenario where the controller and the plant want to share secret messages by using the closed-loop control system operations. Specifically, the controller encodes a message in the control signal $\mathbf{u}(k)$ while the receiver decodes the message from the measured state-space vector $\mathbf{x}(k+1)$. The following proposition proves that for the networked control system in Fig. 2, the eavesdropper channel is a Wyner wiretap-like channel that allows the exchange of secret messages.

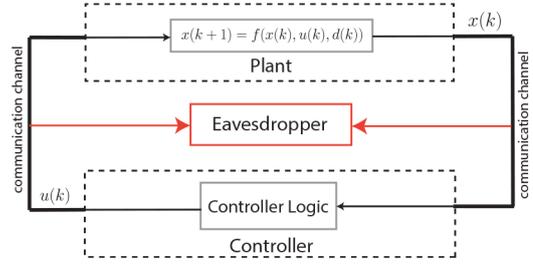

**Fig. 2** *Networked control system*

*Proposition 2:* Consider the networked control system shown in Fig. 2 and the controller and eavesdropper plant models (5) and (10), respectively. Under the condition (11), there exists an encoding scheme allowing the secret exchange of messages between the plant and the controller.

*Proof:* In order to prove the existence of an encoding scheme allowing perfect secrecy, it is sufficient to show that the condition stated in Proposition 1 holds true.

Let $\mathbf{x} = \mathbf{x}(k) \in \mathbb{R}^n$ and $\mathbf{u} = \mathbf{u}(k) = \phi(\mathbf{x}(k))$ be the current plant state vector and command input, respectively, then the following one-step reachable sets (see Definition 2) can be defined

$$\text{Reach}^c(\mathbf{x}, \mathbf{u}) = \{\mathbf{x}^+ \in \mathbb{R}^n : \exists \underline{\mathbf{d}}_c \in \mathcal{D}_c : \mathbf{x}^+ = f_c(\mathbf{x}, \mathbf{u}) + \underline{\mathbf{d}}_c\}$$
$$= f_c(\mathbf{x}, \mathbf{u}) \oplus \mathcal{D}_c \quad (12)$$

$$\text{Reach}^e(\mathbf{x}, \mathbf{u}) = \{\mathbf{x}^+ \in \mathbb{R}^n : \exists \underline{\mathbf{d}}_e \in \mathcal{D}_e : \mathbf{x}^+ = f_c(\mathbf{x}, \mathbf{u}) + \underline{\mathbf{d}}_e\}$$
$$= f_c(\mathbf{x}, \mathbf{u}) \oplus \mathcal{D}_e \quad (13)$$

where $\text{Reach}^c(\mathbf{x}, \mathbf{u})$ and $\text{Reach}^e(\mathbf{x}, \mathbf{u})$ are the one-step reachable sets that can be computed by the controller and eavesdropper, respectively. Given the condition (11), by construction, the following condition holds true:

$$\mathbf{x}(k+1) \in \text{Reach}^c(\mathbf{x}, \mathbf{u}) \subset \text{Reach}^e(\mathbf{x}, \mathbf{u}), \forall (\mathbf{x}, \mathbf{u}) \in \mathbb{R}^n \times \mathbb{R}^m \quad (14)$$

This implies that the eavesdropper has a level of uncertainty (equivocation) about the system one-step evolution $\mathbf{x}(k+1)$ which is always greater than the uncertainty of the closed-loop control system, i.e. $\text{Reach}^c(\mathbf{x}, \mathbf{u}) \subset \text{Reach}^e(\mathbf{x}, \mathbf{u})$. From Proposition 1, the latter is then sufficient to claim that the eavesdropper channel is a Wyner wiretap-like channel. □

### 4.2 Proof-of-concept encoding scheme

In what follows, we show how the available Wyner wiretap-like channel can be leveraged to design an encoding scheme that allows to securely transfer secret messages from the plant to the controller (see Fig. 3). For the sake of clarity, here, we focus our attention on the design of a binary encoding scheme. Nevertheless, the proposed solution can be extended to support a more generic alphabet. Given the uncertain plant model (5), two robustly stabilising state-feedback control policies, namely $\mathbf{u}(k) = \phi_0(\mathbf{x}(k))$ and $\mathbf{u}(k) = \phi_1(\mathbf{x}(k))$, are designed to minimise two different cost indices, namely $J_0$ and $J_1$. Then, the following switching control policy can be defined:

$$\mathbf{u}(k) = \begin{cases} \mathbf{u}_0(k) := \phi_0(\mathbf{x}(k)) & \text{if} \quad b(k) = 0 \\ \mathbf{u}_1(k) := \phi_1(\mathbf{x}(k)) & \text{else} \quad b(k) = 1 \end{cases} \quad (15)$$

where $b(k)$ is an arbitrary binary variable ('0' or '1') that selects which command input ($\mathbf{u}_0(k)$ or $\mathbf{u}_1(k)$) is used at time $k$. For further details, see the *SMTP-CPS* algorithm described at the end of this section where $b(k)$ corresponds to the variable $b_r(k)$ used by the plant (decoder) in the lines 1-2 of the *Plant + Decoder (receiver)* algorithm.

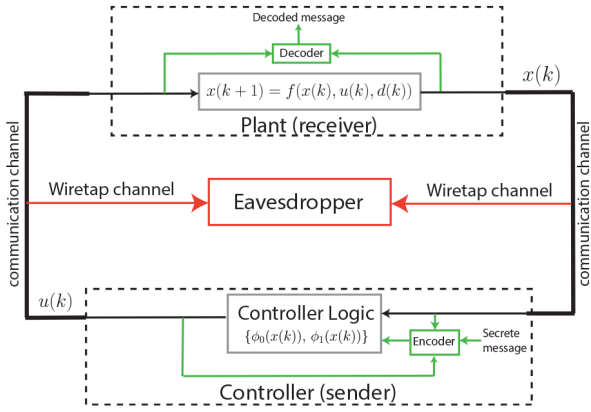

**Fig. 3** *Proposed encoding scheme for networked control systems*

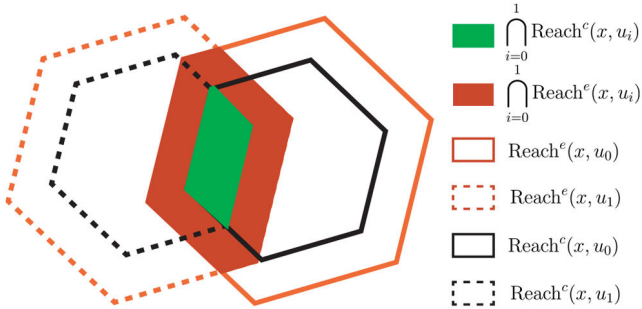

**Fig. 4** *One-step reachable set under the binary encoding scheme: controller versus eavesdropper*

*Remark 2:* Note that the control law $u(k) = \phi_0(x(k))$ and $u(k) = \phi_1(x(k))$ must be properly designed to preserve robust stability under the switching policy (15) (see [39]). For instance, a possible way to achieve this requirement is to design the switching control laws so that a common robust Lyapunov function exists, see e.g. the seminal papers [40, 41] or by resorting to the robust set-theoretic model predictive solution in [42].

By following the Kerckhoffs's principle, we assume that the attacker is aware of the control policy (15).

At each time step $k$, the one-step reachable sets that can be predicted by the controller and attacker are as follows (see Fig. 4):

$$\text{Reach}^c(x(k), u(k)) := \begin{cases} \text{Reach}^c(x(k), u_0(k)) & \text{if} \quad b(k) = 0 \\ \text{Reach}^c(x(k), u_1(k)) & \text{else} \quad b(k) = 1 \end{cases} \quad (16)$$

$$\text{Reach}^e(x(k), u(k)) := \begin{cases} \text{Reach}^e(x(k), u_0(k)) & \text{if} \quad b(k) = 0 \\ \text{Reach}^e(x(k), u_1(k)) & \text{else} \quad b(k) = 1 \end{cases} \quad (17)$$

and the following implications hold true:

$$\exists x^+ \in \bigcap_{i=0}^1 \text{Reach}^e(x, u_i) \text{ such that } x^+ \notin \bigcap_{i=0}^1 \text{Reach}^c(x, u_i) \quad (18)$$

$$\text{If } x^+ \in \bigcap_{i=0}^1 \text{Reach}^c(x, u_i) \text{ then } x^+ \in \bigcap_{i=0}^1 \text{Reach}^e(x, u_i) \quad (19)$$

Therefore, by leveraging the conditions (18) and (19), and by defining as in (20), the one-step reachable set-difference $\text{Diff}(x, u)$

$$\text{Diff}(x, u) := \left( \bigcap_{i=0}^1 \text{Reach}^e(x, u_i) \right) \setminus \left( \bigcap_{i=0}^1 \text{Reach}^c(x, u_i) \right) \quad (20)$$

the following protocol can be defined for the secret exchange of binary vector messages $m$ between the plant and the controller.

In particular, the protocol follows this two phases-procedure, repeated for each secret bit of $m$:

- *First phase*: The controller and the plant agree on a random bit (hereafter denoted as the 'key') that cannot be intercepted by the eavesdropper;
- *Second phase*: The controller uses the previously agreed one-time 'key' to encrypt a single bit of the secrete message.

In the first phase, to agree on a secret bit, the following arguments are exploited. Assume that at the time $k$ the controller computes both $\phi_0(x(k))$ and $\phi_1(x(k))$ and leaves to the plant the decision to apply one of the two actions according to a randomly generated bit, namely $b_r(k) \in \{0, 1\}$. In this scenario, if $x(k + 1) \in \text{Diff}(x(k), u(k))$, then the following implication (arising from (18) and (19)) holds true:

$$x(k+1) \in \text{Diff}(x(k), u(k)) \rightarrow \begin{cases} x(k+1) \in \bigcap_{i=0}^1 \text{Reach}^e(x, u_i) \\ x(k+1) \notin \bigcap_{i=0}^1 \text{Reach}^c(x, u_i) \end{cases} \quad (21)$$

which guarantees that, given $x(k + 1)$, only the controller (but not the eavesdropper) is able to unequivocally determine the value of $b_r(k)$ chosen by the plant. As a consequence, a key is secretly agreed (key = $b_r(k)$), when the condition $x(k + 1) \in \text{Diff}(x(k), u(k))$ is satisfied.

In the second phase, given the agreed upon one-time key, the controller can encrypt the pth bit of $m$, namely $m[p], p \geq 1$, following the information theoretic secure one time pad encryption scheme [19]. In particular, the controller transmits, along with $\phi_0(x(k))$ and $\phi_1(x(k))$, also $b_c(k) = m[p] \veebar \text{key}$, where $\veebar$ denotes the xor operator. In this scenario, the eavesdropper, reading $b_c(k)$ cannot decode $m[p]$ because he/she is not aware of the secret *key* while the plant can simply recover $m[p]$ as $m[p] = \text{key} \veebar b_c(k)$.

Finally, it should be noted that both the plant and controller can determine which phase of the protocol they are executing without any explicit communication channel. The eavesdropper, on the other hand, does not have this knowledge because of the asymmetry in the plant model knowledge available to control system designer and to the eavesdropper.

The algorithm (see Fig. 5) explains the proposed secret message transfer protocol between the plant and the controller.

*Proposition 3:* Consider the networked control system shown in Fig. 3, the controller and eavesdropper plant models (5) and (10), respectively. Under the condition (11), the SMTP-CPS algorithm allows the transfer of secret messages between the controller and the plant (Fig. 6).

*Proof:* To prove the proposition, it is sufficient to collect all the above developments and show that the proposed message exchange protocol satisfies the following properties:

(a) (*Correctness*): The key between the plant and the controller is agreed (see step 4 of the plant and step 1 of the controller) if and only if the controller is certain about the random bit selected by the plant, i.e.

$$x(k+1) \in \text{Reach}^c(x(k), \phi_0(x(k))) \veebar \text{Reach}^c(x(k), \phi_1(x(k))) \quad (22)$$

Therefore, if a secret *key* bit is correctly agreed, then also the encoding/decoding xor operations involving the message bit $m[k]$ and the *key* are correct (see step 11 of the controller and step 8 of the plant) (see [19]).

(b) (*Secrecy*): In this part, we show that the eavesdropper cannot understand the following: when a *key* is agreed, the secret bit message $m[k]$, and the state $s$ of both sender and receiver

*Secret Message Transfer Protocol for CPSs (SMTP-CPS)*

——— *Controller + Encoder (sender)* $\forall k$ ———

**Initialization:** The binary secrete message $\boldsymbol{m}$, the message index $p = 1$, the random bit generator $randBit()$, the initial state of the sender automaton $s = 1$ (Fig. 6a), the set-difference at $k = 0$ $\text{Diff}(\boldsymbol{x}(-1), \boldsymbol{u}(-1)) = \emptyset$.

1: **if** $s == 1$ & $\boldsymbol{x}(k) \in \text{Diff}(\boldsymbol{x}(k-1), \boldsymbol{u}(k-1))$ **then**
2:   $s = 2$;
3:   **if** $\boldsymbol{x}(k) \in \text{Reach}^c(\boldsymbol{x}(k-1), \phi_0(\boldsymbol{x}(k-1)))$ **then** $key = 0$
4:   **else** $key = 1$;
5:   **end if**
6: **else** $s = 1$;
7: **end if**
8: **if** $s == 1$ **then**
9:   $b_c(k) = randBit()$;
10: **else**
11:   $b_c(k) = \boldsymbol{m}[p] \veebar key, p = p + 1$;
12: **end if**
13: Send $[\phi_0(\boldsymbol{x}(k)), \phi_1(\boldsymbol{x}(k)), b_c(k)]$ to the plant;

——— *Plant + Decoder (receiver)* $\forall k$ ———

**Initialization:** The binary secrete message $\boldsymbol{m} = [\,]$, the message index $p = 1$, the random bit generator $randBit()$, the initial state of the receiver automaton $s = 1$ (Fig. 6b).

1: $b_r(k) = randBit()$.
2: Apply $\boldsymbol{u}(k) = \phi_{b_r(k)}(\boldsymbol{x}(k))$;
3: **if** $s == 1$ **then**
4:   **if** $\boldsymbol{x}(k+1) \in \text{Diff}(\boldsymbol{x}(k), \boldsymbol{u}(k))$ **then**
5:     $key = b_r(k), s = 2$;
6:   **end if**
7: **else**
8:   $\boldsymbol{m}[p] = key \veebar b_c(k), p = p + 1, s = 1$;
9: **end if**

**Fig. 5** *Algorithm: secret message transfer protocol for CPSs (SMTP-CPS)*

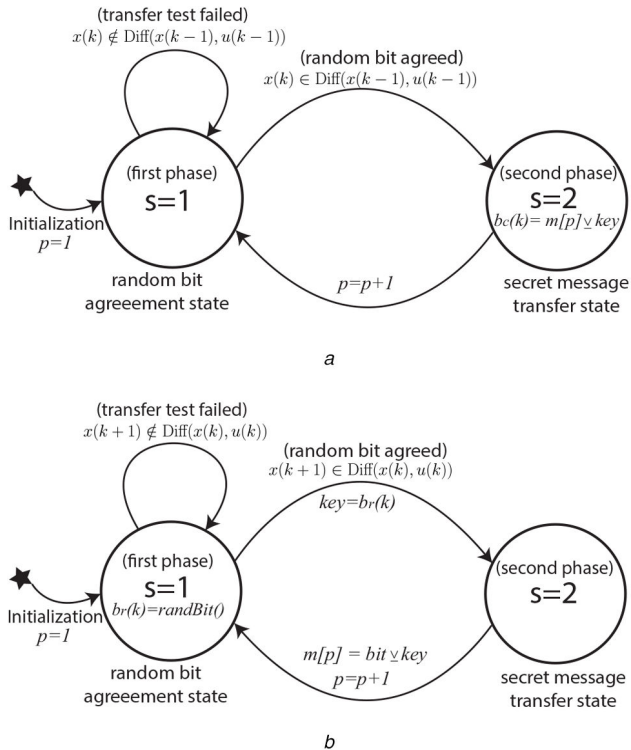

**Fig. 6** *Secrete message transfer automata for*
*(a)* Sender, *(b)* Receiver

automaton. In particular, given the attacker resources, the eavesdropper can compute the attacker reachable sets (17) but the eavesdropper cannot calculate the controller reachable set (16). As a consequence, the attacker cannot determine the set difference (20) and understand when the *key* agreement condition $\boldsymbol{x}(k + 1) \in \text{Diff}(\boldsymbol{x}(k), \boldsymbol{u}_i(k))$ is satisfied. The latter is sufficient to conclude that the eavesdropper is not aware of the *key* of the secret bit messages $\boldsymbol{m}[k]$ and the state $s$ of the automaton, concluding the proof.□

*Remark 3:* The computational overhead associated with the proposed secret message transfer protocol is dominated by the computation of:

- The reachable sets $\text{Reach}^c(\boldsymbol{x}(k), \boldsymbol{u}_i)$ and $\text{Reach}^e(\boldsymbol{x}(k), \boldsymbol{u}_i)$, $i = 1, 2$.
- The set membership tests $\boldsymbol{x}(k) \in \text{Diff}(\boldsymbol{x}(k-1), \boldsymbol{u}(k-1))$ and $\boldsymbol{x}(k) \in \text{Reach}^c(\boldsymbol{x}(k-1), \boldsymbol{u}_0)$

According to (12) and (13), the computations of the reachable sets are obtained by performing Minkowski sums between the vector $f_c(\boldsymbol{x}, \boldsymbol{u}_i)$ and the disturbance sets $\mathcal{D}_c$ and $\mathcal{D}_e$, respectively. By assuming a polytopic convex representation of the disturbance sets, such operations have a computational complexity $\mathcal{O}(n_v^{\lfloor n/2 \rfloor})$ that depends on the dimension $n$ of the state-space vector $\boldsymbol{x}(k)$ and the number of vertices $n_v$ needed to describe $\mathcal{D}_c$ and $\mathcal{D}_e$ (see [43–46] and references therein).

As for the computational complexity of the set membership tests, it should be noted that the set-difference $\text{Diff}(\boldsymbol{x}(k-1), \boldsymbol{u}(k-1))$ does not need to be explicitly computed. Indeed, as shown in (21), the set membership test $\boldsymbol{x}(k) \in \text{Diff}(\boldsymbol{x}(k-1), \boldsymbol{u}(k-1))$ is equivalent to simple logic condition on the result of the following set-membership tests:

$$\boldsymbol{x}(k) \in \text{Reach}^c(\boldsymbol{x}(k), \boldsymbol{u}_i), \quad \boldsymbol{x}(k) \in \text{Reach}^e(\boldsymbol{x}(k), \boldsymbol{u}_i), \ i = 1, 2$$

Also, if the reachable convex sets are presented as polytopes, then each test is equivalent to solving a simple linear programming (LP) problem whose computational complexity is polynomial in time [47]. Moreover, if the number of optimisation variables is small (the size of the state vector $\boldsymbol{x}(k)$ in the specific case), then the complexity of solving the LP problem becomes linear in time w.r.t. to the number of constraints (inequalities) [48].

Finally, it should be noted that for systems with a large number of state variables, the complexity of the encoding/decoding scheme can be reduced by simply implementing the proposed algorithm on a small subset of the state-space variables.

## 5 Simulation results

In this section, the effectiveness of the proposed Wyner wiretap-like encoding scheme is validated by means of a numerical simulation example. The simulation has been implemented in Matlab where the MPT3 toolbox [49] has been used to design the proposed controller and compute the one-step reachable sets (16) and (17) required to implement the SMTP-CPS algorithm.

The plant dynamics (3) are defined by a discrete linear time-invariant system subject to additive exogenous bounded disturbances $\boldsymbol{x}(k + 1) = \boldsymbol{A}\boldsymbol{x}(k) + \boldsymbol{B}\boldsymbol{u}(k) + \boldsymbol{d}(k)$, whose sampling time is $T = 0.1\,s$, the system matrices are

$$\boldsymbol{A} = \begin{bmatrix} 1 & 0.0975 \\ 0 & 0.9512 \end{bmatrix}, \quad \boldsymbol{B} = \begin{bmatrix} 0.0246 \\ 0.4877 \end{bmatrix}$$

and the disturbance $\boldsymbol{d}(k)$ is described by the following component-wise bounds:

$$-0.1 \leq d_j(k) \leq 0.1, \quad j = 1, 2 \tag{23}$$

Moreover, the input saturation constraint $\boldsymbol{u}(k) \in \mathcal{U} = \{\boldsymbol{u} \in \mathbb{R}^2 : |\boldsymbol{u}| \leq 6\}$ is imposed.

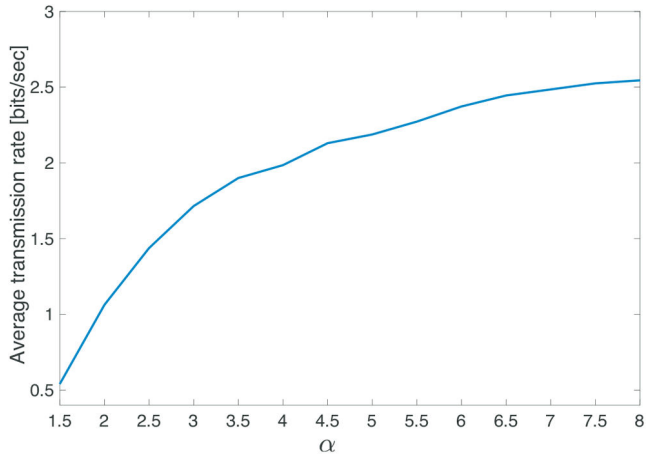

**Fig. 7** *Secrete average message transfer rate versus $\alpha (\mathcal{D}_e = \alpha \times \mathcal{D}_c)$*

We assume that the identified plant model (5) is described by the same uncertain dynamical model $x(k+1) = Ax(k) + Bu(k) + d_c(k)$, with a disturbance $\mathcal{D}_c$ defined by the following increased component-wise bounds:

$$-0.12 \leq d_{c_j}(k) \leq 0.12, \quad j = 1, 2 \quad (24)$$

Given the uncertain identified plant model, the set-theoretic model predictive control scheme proposed in [42] has been used to implement the switching control law (15). For the sake of clarity and completeness, first, the controller design steps are summarised, then the switching control laws $\phi_0(k)$ and $\phi_1(k)$ are presented. The reader is referred to [42] for further details.

### 5.1 Controller design steps

(i) A linear quadratic state-feedback controller $u(k) = Kx(k)$ has been computed using the capabilities of the MPT3 toolbox, and the smallest terminal robust positively invariant region $\mathcal{T}_0$, associated to the controller, has been determined [50]. The computed controller gain vector $K$ is

$$K = [-13.27 \quad -2.26] \quad (25)$$

(ii) The terminal region $\mathcal{T}_0$ is then enlarged to compute the set of states that can be steered into $\mathcal{T}_0$ in a finite number, $N > 0$, of steps. The latter has been obtained by recursively applying the following definition of robust one-step controllable set (see [36]):

$$\mathcal{T}_i := \{x \in \mathbb{R}^n : \forall d_c \in \mathcal{D}_c, \exists u \in \mathcal{U} : Ax + Bu + d_c \in \tilde{\mathcal{T}}_{i-1}\} \quad (26)$$

where $\tilde{\mathcal{T}}_{i-1} := \mathcal{T}_{i-1} \sim \mathcal{D}_c$ and $\sim$ denotes the Minkowski set difference operator [35]. For our simulations, we set $N = 250$.

### 5.2 Control action computation

Given the offline computed family of one-step controllable sets $\{\mathcal{T}_i\}_{i=0}^{250}$ and a feasible initial condition $x(0) \in \bigcup_{i=0}^{250} \mathcal{T}^i$, at each sampling time instant $k$, the control action $u(k)$ is computed as follows:

- Let $i(k) := \min\{i : x(k) \in \mathcal{T}_i\}$
- If $i(t) == 0$ then

$$u(t) = Kx(t),$$

Else solve the following quadratic optimisation problem:

$$\begin{aligned} u(k) = \arg\min_u J(k) \quad \text{s.t.} \\ Ax(k) + Bu \in \tilde{\mathcal{T}}_{i(k)-1}, u \in \mathcal{U} \end{aligned} \quad (27)$$

where $J(k)$ is any the convex cost function of interest.

In [42], it is shown that the cost function $J(k)$ can be arbitrary chosen at each sampling time without compromising the system stability or convergence to the terminal set $\mathcal{T}_0$.

Throughout our simulations, we have associated the switching control laws $\phi_0(x(k))$ and $\phi_1(x(k))$ to the following convex cost functions:

$$J_0(k) = \|Ax(k) + Bu(k)\|_2^2, \quad J_1(k) = \|u(k)\|_2^2$$

which minimise the time to reach the terminal region and the control effort, respectively. We have evaluated the performance of the proposed *SMTP-CPS* encoding scheme (in terms of bits/s) for different attacker's disturbance sets $\mathcal{D}_e \supset \mathcal{D}_c$. In particular, we have considered eight different attacker's disturbance sets $\mathcal{D}_e$ such that $\mathcal{D}_e = \alpha \times \mathcal{D}_c$, where $\alpha \in \{1.5, 2, ..., 7.5, 8\}$. Moreover, we have considered three different plant's initial conditions, and each simulation has been run for 50 time steps (5 s) and repeated 20 times to take into account different disturbance $d(k)$ realisations.

The simulation results are reported in Fig. 7 where $\alpha$ is depicted on the *x-axis* and the average obtained transmission rates (bits/s) are shown on the *y-axis*. It is clear that increasing the ratio between the eavesdropper and controller disturbance set, then the transmission rate increases. The latter finds justification in the fact that $\alpha$ quantifies the model asymmetry between the controller (defender) and the eavesdropper.

## 6 Conclusions and open research problems

By exploiting asymmetry in the plant model knowledge available to control system (the defender) and to the eavesdropper, we showed the existence of Wyner wiretap-like encoding scheme for CPSs. This asymmetry plays a role similar to the one performed by the noise in the original Wyner's wiretap communication channel. Such scheme allows us to solve the problem of exchanging secrete messages/keys in CPSs without resorting to traditional cryptographic solutions. Up to the authors' knowledge, no prior work has considered the utilisation of the underlying dynamics of CPSs to create such a secure channel between the networked controller and the plant. One of the main advantages of our scheme is that its security does not rely on any assumptions about the computational power of the adversary. In other words, unlike symmetric key and most public-key cryptographic solutions, our proposed scheme achieves its claimed security properties even if the adversary has unlimited computational power.

For future works, open research problems include: (i) quantifying the theoretical secrecy capacity of the investigated CPS wiretap channels in terms of $\mathcal{D}_c$ and $\mathcal{D}_e$, (ii) extending the results to different classes of robust or stochastic control models. For example, throughout our analysis, we assumed that the uncertainty sets $\mathcal{D}_c$ and $\mathcal{D}_e$ are bounded and uniformly distributed. However, in general, other distributions (e.g. Gaussian, truncated Gaussian) might better suit the underlying CPS system, (iii) determining the optimal source alphabet size for a given CPS setup, (iv) designing of a CPS encoding techniques that can approach the fundamental secrecy capacity limits in practice.